\documentclass[aps,pre,twocolumn,showpacs,preprintnumbers,amsmath,amssymb]{revtex4}

\usepackage{graphicx}
\usepackage{dcolumn}

\newcommand{\vect}[1]{{\mbox{\boldmath $#1$}}}
\newcommand{\bm}[1]{\vect{#1}}
\newcommand{\Sec}[1]{Sec.~\ref{sec:#1}}
\newcommand{\App}[1]{Appendix~\ref{sec:#1}}
\newcommand{\Eq}[1]{(\ref{#1})}
\newcommand{\Fig}[1]{Fig.\ref{fig:#1}}
\newcommand{\Tab}[1]{Table~\ref{tab:#1}}

\newcommand{\deffig}[4]{
\begin{figure}[tb]
  \begin{center}
  \includegraphics[width=#3 \textwidth]{#2}
  \end{center}
  \caption{ \label{fig:#1} #4}
\end{figure}
}

\begin{document}

\title{
  Coarse-grained loop algorithms for Monte Carlo simulation of quantum 
  spin systems
}

\author{
  Kenji~Harada$^1$ and Naoki~Kawashima$^2$
}
\affiliation{
  ${}^1$ Department of Applied Analysis and Complex Dynamical Systems, 
  Kyoto University, Kyoto 606-8501, Japan\\
  ${}^2$ Department of Physics, Tokyo Metropolitan University, 
  Tokyo 192-0397, Japan\\
}

\date{28 October 2002}

\begin{abstract}
Recently, Sylju\r{a}sen and Sandvik proposed a new 
framework for constructing algorithms of quantum Monte Carlo 
simulation. While it includes new classes of powerful algorithms, 
it is not straightforward to find an efficient algorithm 
for a given model. 
Based on their framework, we propose an algorithm
that is a natural extension of the conventional loop algorithm
with the split-spin representation.
A complete table of the vertex density and the worm-scattering probability
is presented for the general $XXZ$ model of an arbitrary $S$
with a uniform magnetic field.
\end{abstract}

\pacs{
  02.70.Ss, 02.70.Uu, 05.10.Ln, 75.40.Mg
}

\maketitle

\section{Introduction}
Among many numerical techniques for condensed matter physics, 
the Monte Carlo method is a popular choice when a long correlation 
length or a small excitation gap is anticipated. 
Apart from the negative sign difficulty, most of the shortcomings
of the quantum Monte Carlo (QMC) method for spin systems have been 
removed or reduced.
In particular, the QMC for finite temperature
based on the path-integral representation has been improved 
considerably during the past decade.
The improvement was achieved mainly by the development of the
loop-cluster algorithms\cite{LoopClusterQMC} and related methods.
For instance, the critical slowing down was tamed by the
loop-cluster algorithms for a broad class of quantum 
spin systems\cite{HaradaKXY}.
It was shown\cite{Wolff} that the typical size of the clusters coincides with
the correlation length.
Because of this property, an effective update of configurations
is possible.
The other slowing down, due to small intervals for discretization
of the imaginary time, was completely removed also by the loop-cluster
QMC\cite{KawashimaG1995}, 
which became even more evident by the extension to
continuous imaginary time\cite{BeardW1996}.
An efficient measurement of some of important off-diagonal quantities 
was made possible through the improved estimator\cite{Broweretal}.

One of the difficulties that have been left unsolved until recently
was the freezing of configurations due to an external field competing
with the exchange couplings.
In the conventional framework of the loop-cluster algorithms, 
the field term does not affect the graph assignment probabilities.
It is taken into account only in the flipping probabilities of clusters.
As a result, the cluster size does not correspond to the physical
correlation length any more.
It was demonstrated\cite{Worm}
that this difficulty can be removed,
in the case of the $S=1/2$ antiferromagnetic Heisenberg model,
by introducing two singular points at which the local conservation
rule of particle number (or magnetization) is violated.
These singular points are called ``worms.''
This extension of the configuration space makes it possible to
take the external field into account in the hopping
probability of worms.

Another difficulty is large memory requirement due to
the split-spin representation\cite{KawashimaG1995}.
When one uses the loop algorithm for a spin problem with large $S$,
it is customary to replace each spin operator 
by a sum of $2S$ Pauli matrices.
Therefore, for larger $S$, the algorithm consumes more memory.
The stochastic series expansion (SSE)\cite{SSE,SSEreview} does not have this
difficulty, since it works directly on the original spin-configuration space.
The SSE is based on the high-temperature series expansion of the partition 
function, rather than the path-integral formulation.
However, it was pointed out\cite{Troyer2001,SyljuasenS2002} that these
two apparently different formulations are essentially equivalent
in the limit of the infinite order expansion.
The apparent difference was due to the different
updating method, rather than the formulations themselves.

Quite recently\cite{SyljuasenS2002}, Sylju\r{a}sen and Sandvik 
introduced the notion of ``directed loops'' and
proposed a framework that accommodates all of the above-mentioned
ideas, i.e., the loop updating, the worm updating, and
the two formulations.
Their framework can be compared with Kandel and Domany's framework
\cite{KandelD1991} for the loop-cluster algorithms.
In fact, the mathematical formulation of the Sylju\r{a}sen-Sandvik (SS)
scheme has a similar structure to Kandel and Domany's 
(see \App{Framework}),
and the resulting algorithm coincides with the loop-cluster algorithm in some cases.
In this sense, the SS scheme can be viewed as a generalization of
the Kandel-Domany framework.

An algorithm based on the SS scheme
is characterized by the scattering probabilities of worms.
Although the detailed balance condition imposes a set of equations
to be satisfied by these probabilities, there are still a lot of
degrees of freedom.
Some of the solutions to these detailed balance equations lead
to the single-cluster version of the conventional 
loop-cluster algorithm at zero magnetic field, 
which are known to be efficient.
However, all the solutions are not necessarily efficient or practical.
There are obviously many bad solutions in which the ``back-tracking'' probability
\cite{SyljuasenS2002} or ``turning-back'' probability are dominating.
In addition, the straightforward solutions of the heat-bath type
do not work either, as we see below in \Sec{Performance}.
Although an efficient solution was discussed\cite{SyljuasenS2002} for $S=1/2$
$XXZ$ models, the prescription was not given for general $S$.

Similar to the Kandel-Domany framework, the SS scheme does not
give a concrete prescription for obtaining a good solution
that leads to an efficient algorithm for specific models.
A rule of thumb for obtaining a good solution is 
to minimize the turning-back probability.
However, even if the turning-back probability is fixed,
we still have many degrees of freedom to play with, and
the efficiency of the algorithm strongly depends on the choice of
the worm-scattering probabilities, as we demonstrate in \Sec{Performance}.
While this freedom can be quite useful for constructing
new types of efficient algorithms,
it makes finding a reasonable solution a nontrivial task.

In this paper, we propose a natural extension of existing algorithms
that determines a unique set of scattering probabilities of worms.
The resulting algorithm is within the SS scheme and
expected to be efficient for a wide class of quantum spin systems.
The algorithm can be obtained by the coarse graining mapping applied
to an algorithm in the split-spin representation.
In \Sec{SplitSpin}, we first discuss algorithms in the split-spin 
representation.
In \Sec{CoarseGraining}, we show how we can obtain an algorithm 
in the original spin representation by coarse-graining the 
split-spin algorithm.
Based on this general prescription for the coarse-grained algorithms,
we present in \Sec{XXZ}
a complete table of the worm scattering probability
for the $XXZ$ models with arbitrary $S$.
Finally, in \Sec{Performance}, we compare the present algorithm with 
other algorithms such as the directed loop algorithm with the
straightforward heat-bath solution.

\section{Algorithms in the split-spin representation}
\label{sec:SplitSpin}
It has been pointed out in previous papers\cite{Troyer2001,SyljuasenS2002} that,
in the limit of the infinite order expansion,
a Monte Carlo algorithm based on the series expansion
can be reformulated in terms of the language of the path-integral 
with continuous imaginary time, and vice versa.
In what follows, therefore, we describe algorithms in this limit 
for the sake of simplicity, and use the path-integral language.
The translation into the series-expansion language and
its modification for a finite order expansion should be straightforward.

A simulation based on the path-integral representation (or the SSE in
the infinite order expansion limit) can be visualized in a $(d+1)$ -
dimensional space-time where $d$ is the real-space dimension.
At each point in this space-time an integral variable is defined, 
and it takes on one of the $2S+1$ values $-S, -S+1, \dots,$ and $S$.
In the case of $S=1/2$ the variables are one-bit (or Ising) variables.
Accordingly, we consider world lines, which are trajectories
of up-spins in this space-time.
In the present paper, we use the term 
``world-line configurations'' to refer to the 
spin configurations in the space-time for general $S$,
although they are not represented by simple lines for $S>1/2$.
A Monte Carlo algorithm is nothing but a procedure by which 
the world-line configuration is updated so that 
the limiting probability distribution may coincide with
the weight of the configuration, 
i.e., the exponential of the action.

In the SS scheme, we deal with objects defined in the $(d+1)$-
dimensional space-time (\Fig{Objects}).
\deffig{Objects}{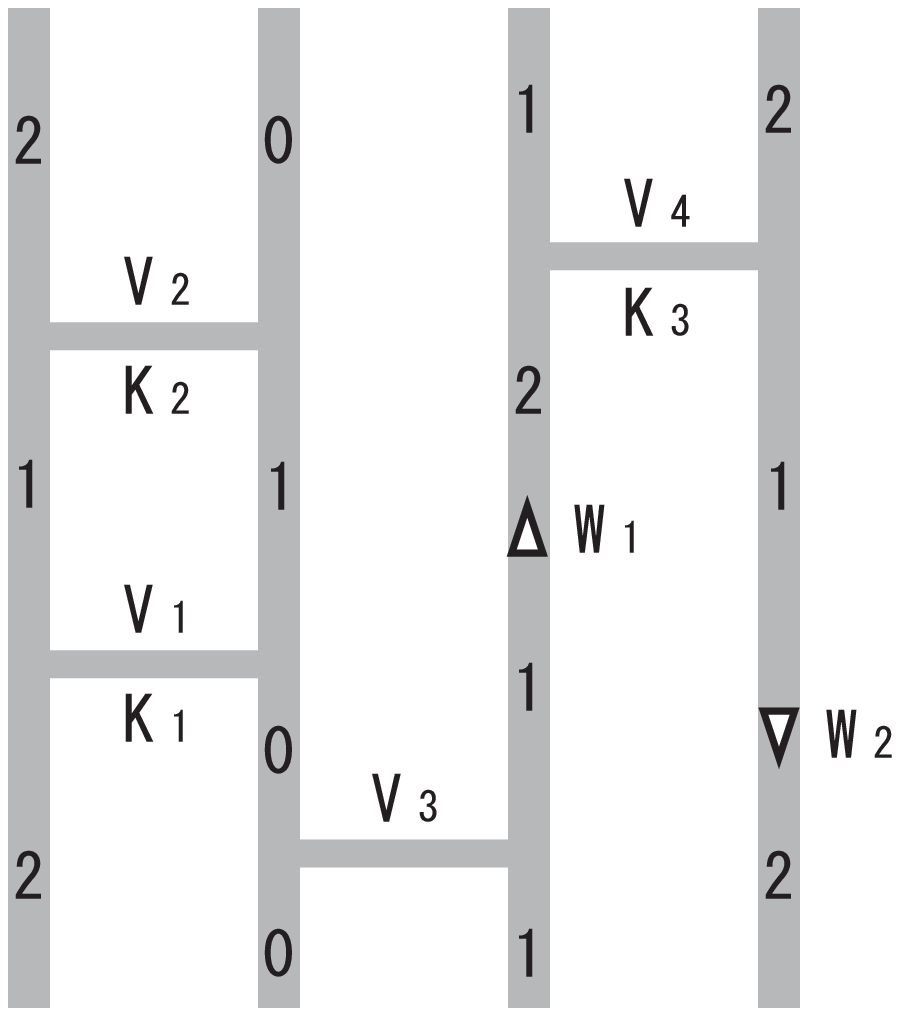}{0.28}{
Various objects in $(1+1)$-dimensional space-time
in the case of $S=1$.
The vertical direction corresponds to the temporal coordinate
whereas the horizontal direction corresponds to the spatial coordinate.
Kinks ($K_1, K_2$, and $K_3$),
vertices ($V_1, V_2, V_3$, and $V_4$),
and worms ($W_1$ and $W_2$) are shown.
A number ($0,1$, or $2$) printed on every segment
means the number of particles [$\equiv S^z_i(\tau) + S$].
Horizontal lines represent vertices.
The kinks coincide with the vertices at which 
particles jump from segments to segments.
}
A vertical line of length $\beta$ represents a spin.
A {\it kink} is a point at which the local spin configuration
changes.
A particle (or an up-spin) jumps from one vertical 
line to another only at kinks.
In models without particle number (or magnetization) conservation,
a point at which a particle disappears or appears is also a kink.
Every kink is located on a {\it vertex}.
Vertices in the SS scheme play a role comparable to that of
local graph elements in the conventional loop-cluster algorithms.
In particular, for models in which the magnetization conserves,
vertices are represented by short horizontal lines,
each connecting two or more neighboring vertical lines.
If a vertex connects two lines, we call such a vertex {\it four legged} 
since it joins four {\it segments}, where a segment is 
a part of a vertical line which is delimited by two vertices.
A {\it worm} is a kink of a special kind located on a segment
\cite{DefinitionOfWorm}.
A worm can move continuously as the simulation proceeds,
while locations of ordinary kinks and vertices are fixed 
until they are deleted.
In the applications discussed in the present paper,
there are only zero or two worms at the same time 
in the whole system.

For quantum spin systems, 
one cycle of update in the SS scheme consists of the following
operations on these objects:
1) assigning vertices to a given world-line configuration,
2) creating a pair of worms,
3) letting one of them move along segments and be scattered by vertices
   until it comes back to the other worm to be annihilated, and
4) deleting all the vertices with no kinks on them.
In the rest of the present paper, we see these operations in more detail.

When a world-line configuration is given, we first assign vertices.
Vertices are assigned to every part of the system probabilistically
with a density that depends on the local world-line configuration.
In addition, all the kinks are regarded as vertices.
After placing all vertices, we choose a point on a segment 
at random and create a pair of worms there.
Then, one of the worms starts moving.
As a worm passes a point in the space-time,
it changes the local spin value there.
When the moving worm encounters a vertex, it may be scattered.
The outgoing direction after the scattering is determined
stochastically with certain predetermined scattering probabilities.
When a moving worm meets the other worm, they annihilate.
Therefore, what we have to specify in order to define an algorithm are 
the density of vertices and
the scattering probability of worms.
The SS scheme imposes conditions on these.
The conditions are summarized in \App{Framework}.

When spins in a given model are larger than S=1/2,
it is customary to replace the original spin operators 
by the sum of $2S$ Pauli spins\cite{KawashimaG1995,HaradaTKXY}, i.e.,
$$
  S^{\alpha}_i \rightarrow \tilde S^{\alpha}_i \equiv
  \sum_{\mu=1}^{2S} \sigma^{\alpha}_{i\mu}
  \qquad (\alpha = x,y,z),
$$
where $\sigma^{\alpha}_{i\mu}$ is an $S=1/2$ spin operator.
The partition function is expressed in terms of these
$\sigma$ degrees of freedom.
Since the original phase space corresponds to the subspace in which
$$
  (\tilde S^x_i)^2 + (\tilde S^y_i)^2 + (\tilde S^z_i)^2 = S(S+1)
$$
holds, we have to project out all the states orthogonal 
to this subspace to obtain the correct partition function.
This can be done by inserting the projection operator $P$:
$$
  Z 
  = {\rm Tr}_{\{S_i\}} (e^{-\beta H[\{S_i\}]})
  = {\rm Tr}_{\{\sigma_{i\mu}\}} (P e^{-\beta H[\{\tilde S_i\}]}).
$$
Hereafter, the representation based on $S_i$ degrees of freedom is 
referred to as
the ``original spin representation'' whereas that based on $\sigma_{i\mu}$
the ``split-spin representation.''

For many models, it is rather straightforward to obtain 
an algorithm in the split-spin representation.
For example, we can obtain an algorithm for 
the $XXZ$ models with arbitrary magnitude of spins $S$
from that for the corresponding model with $S=1/2$,
simply by regarding the former as
a superposition of many of the latter.
If we do so, we consider $2S$ vertical lines for each original spin.
Accordingly, a point in the space-time is specified by 
three numbers $((i,\mu),\tau)$ rather than two $(i,\tau)$.
The coupling between two original spins $S_i$ and $S_j$
is transformed into $(2S)^2$ couplings, each couples $\sigma_{i,\mu}$ 
and $\sigma_{j,\nu}$:
\begin{eqnarray*}
  -\tilde H_{ij} 
  & \equiv &
    J \tilde S^x_i \tilde S^x_j 
  + J \tilde S^y_i \tilde S^y_j 
  + J' \tilde S^z_i \tilde S^z_j 
  + \frac{H_p}{2}(\tilde S^z_i + \tilde S^z_j), \\
  & = & 
    \sum_{\mu,\nu} \left( \rule{0mm}{5mm}
    J \sigma^x_{i\mu} \sigma^x_{j\nu} 
  + J \sigma^y_{i\mu} \sigma^y_{j\nu} 
  + J' \sigma^z_{i\mu} \sigma^z_{j\nu} \right. \\
  & & \qquad\qquad\qquad\qquad\qquad \left. \rule{0mm}{5mm}
  + \frac{h}{2}(\sigma^z_{i\mu} + \sigma^z_{j\nu}) \right),
  \label{XYZ}
\end{eqnarray*}
where $J>0$, $h \equiv H_p/(2S)$, and $H_p$ is the external field per bond
(e.g., $H_p = H/d$ for the hypercubic lattice where $H$ is 
the external field per site).
For the vertex assignment, we apply the procedure for 
the directed loop algorithm for $S=1/2$\cite{SyljuasenS2002}
to every one of $(2S)^2$ combinations of $\sigma$ spins.
To be more precise, the density for the vertex between
two split spins is the same as that in the directed loop algorithm
for $S=1/2$ with $H_p$ replaced by $h\equiv H_p/(2S)$.
Similarly, the worm-scattering probabilities for $S=1/2$ can be used
for split spins with the same modification of $H_p$.

For the projection operator, we do essentially the same as we usually
do in  the conventional loop algorithm for $S>1/2$\cite{HaradaTKXY}.
In the present framework, we represent it by special vertices,
each located at $\tau=\beta$ connecting all the $(2S)$ vertical lines 
on a site $i$.
To be specific, when a worm moves upwards along the vertical line $(i,\mu)$ 
and hits the point $((i,\mu),\beta)$ from below,
it jumps to $((i,\nu),0)$ and go on upwards.
The line to which the worm jumps, $(i,\nu)$, is chosen with equal 
probability among those on which the local spin state is the same as
the spin state right above the incoming worm.
Namely, it is chosen among such $\nu$'s that
$\sigma_{i\nu}(0) = \sigma_{i\mu}(\beta)$ may hold.

\section{Coarse Graining}
\label{sec:CoarseGraining}
One of the drawbacks of the split-spin representation mentioned 
above is that it may require much more memory than the original
spin representation.
For example, in the loop algorithm for the $SU(N)$ models
\cite{HaradaKT2002},
we insert graphs that involve all Pauli spins on two neighboring sites
at the same imaginary time (\Fig{SUN}).
\deffig{SUN}{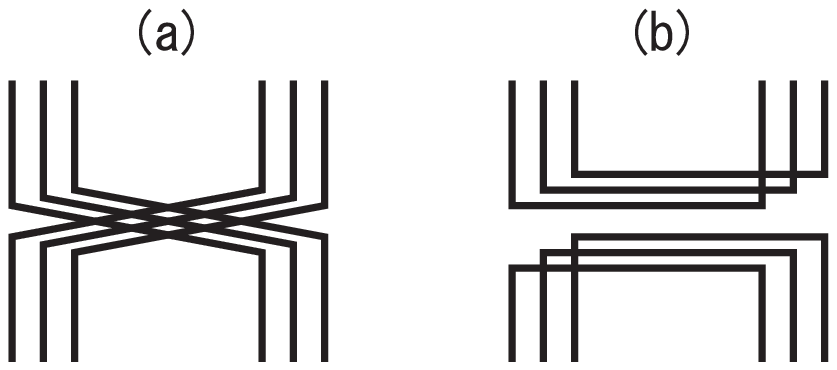}{0.4}{
Two types of 12-legged vertices that appear in the $S=3/2$ SU(4) models 
in the split-spin representation.
}
In the split-spin representation, insertion of a graph of this type
creates $2(N-1) = 4S$ new segments.
Since the memory requirement is roughly proportional to the number of
segments, a loop algorithm for the $SU(N)$ model requires memory
resources proportional to $2(N-1)$.
If we can construct an algorithm in the original spin representation,
insertion of a graph would create only two new segments.
This leads to a memory requirement smaller by factor $1/(N-1)$
than that of the split-spin representation.

Another drawback is the lack of portability of the code.
In the split-spin representation, there are many kinds of vertices,
in principle, depending on the number of legs.
Therefore, we have to change the core part of the code 
to accommodate new kinds of graphs for each model
unless we implement all possible sorts of graphs initially,
which is impractical.
On the other hand, in the coarse-grained representation, 
all the vertices are four-legged 
(for models with two-body interactions)
and there are only four different types of scattering of worms.
Therefore, the core part of codes for all models are 
the same except for the densities of vertices and
the scattering probabilities of worms.
For example, if we have a code based on the SS scheme for
the $SU(N)$ model we can immediately obtain a code 
for the $XY$ model simply by changing the arrays of
the probability tables.

In order to take full advantage of the SS scheme, therefore,
we have to construct probability tables for algorithms based on
the original spin representation rather than the split-spin representation.
For this purpose, we consider a ``coarse-graining'' map 
and its stochastic inverse.
The map is basically disregarding the detailed information of split spins.
The inverse of the map is to choose stochastically one of the split-spin 
configurations which are transformed by coarse graining
into a given original-spin configuration.
With these maps and an algorithm in the split-spin representation,
we can construct an algorithm in which we manipulate only original 
spin degrees of freedom.

To illustrate the idea in more detail, we again take the $XXZ$ model 
with an arbitrary magnitude of spins $S$.
We can define a coarse-graining map from a split-spin world-line
configuration $\tilde{\bm{S}}$ into an original spin world-line 
configuration $\bm{S}$ as
$$
  \tilde{\bm{S}} \equiv \{\sigma_{i\mu}(\tau)\}\rightarrow 
  \bm{S} \equiv \{S_{i}(\tau)\},
$$
where $\sigma_{i\mu}(\tau)$ is the value of $\sigma^z_{i\mu}$ at the
imaginary time $\tau$, whereas $S_{i}(\tau)$ is the sum of them, i.e.,
$S_{i}(\tau) \equiv \sum_{\mu} \sigma_{i\mu}(\tau)$.
Similarly, we can define a map for vertices.
Since the only interaction is of second order in the spin operators,
a vertex of the $XXZ$ model has only four legs (i.e., connects only two lines).
If a vertex connects two lines $(i,\mu)$ and $(j,\nu)$, 
we associate with it a vertex that connects two coarse-grained lines 
$i$ and $j$.
Of course, in the latter representation, the information concerning
the indices $\mu$ and $\nu$ is missing.

Obviously, these mappings are many-to-one mappings.
However, we can define the inverse of this coarse-graining map.
In the following, we adopt the ``particle'' picture in which
an up-spin is regarded as a particle whereas a down-spin a hole.
Correspondingly, we use particle numbers $l_i(\tau) = 0, 1,
\dots,2S$ and $n_{i\mu}(\tau)=0,1$, 
instead of $S_i(\tau)$ and $\sigma_{i\mu}(\tau)$,
to specify local states of spins.
These are related to $S_i(\tau)$ and $\sigma_{i\mu}(\tau)$ by
$S_i(\tau) = l_i(\tau)-S$
and $\sigma_{i\mu}(\tau) = n_{i\mu}(\tau)-1/2$.

The inverse mapping of a local state is rather simple.
Suppose that a model is an $S=1$ model and
a local spin state at the point of interest is $l_i(\tau) = 1$ 
in the coarse-grained representation.
There are two split-spin states that are mapped to this state,
i.e., $(n_{i1}(\tau), n_{i2}(\tau))=(1,0)$ and $(0,1)$.
Both configurations are chosen with the same probability (i.e., 1/2)
since there is no reason to put any bias.
For general $S$, all configurations that satisfy $\sum_\mu n_{i\mu}=l$
are chosen with the same probability,
where $l=0,1,\dots,2S$ is the local state on the coarse-grained line.

The inverse mapping of a vertex can be defined in a similar way.
When two space-time points $(i,\tau)$ and $(j,\tau)$ 
are connected by a vertex (with no kink on it)
in the coarse-grained representation,
we can map it to a vertex connecting $((i,\mu),\tau)$ and 
$((j,\nu),\tau)$ with some probability.
When $S=1$, there are four different ways of choosing 
$\mu$ and $\nu$ that are to be connected.
However, the probability for taking one of them is not 1/4 in this case.
This is because the density of vertices depends on the 
spin states at their legs.
If, for example, the density of vertex between two particles
is higher than that between a particle and a hole,
a given coarse-grained vertex should be mapped to the former with larger
probability than the latter.
In other words, the probability for associating a coarse-grained vertex
with a particular split-spin vertex should be proportional to the
density of the latter.
Therefore, the probability for associating a coarse-grained
vertex between $i$ and $j$ with a split-spin vertex between 
$(i,\mu)$ and $(j,\nu)$ is given by
\begin{equation}
  \frac{\rho^{{\rm (ss)}}_{n_{i\mu}n_{j\nu}}}
  {\sum_{\mu,\nu}\rho^{{\rm (ss)}}_{n_{i\mu}n_{j\nu}}}.
  \label{DensityOfVertices}
\end{equation}
Here, $\rho^{{\rm (ss)}}_{nn'}$ is the density of vertices
in the split-spin representation
where the local spin values are $n$ and $n'$ at the 
legs of vertices. 

The coarse-graining map and its inverse can be used for obtaining
the vertex density and the worm-scattering probability
in the coarse-grained representation
from those in the split-spin representation.
Suppose an imaginary-time interval in which the state of
two neighboring sites $i$ and $j$ are specified by $l$ and $m$, respectively.
The site $i$ consists of $l$ particles and $\bar l$ ($\equiv 2S-l$) holes
whereas the site $j$ consists of $m$ particles and $\bar m$ holes.
Then, there are $lm, l\bar m, \bar l m$, and $\bar l \bar m$
possible combinations of $11$, $10$, $01$, and $00$ pairs of
$\sigma$ spins, respectively.
Since we assign a vertex with density $\rho^{{\rm (ss)}}_{11}$
for each $11$ pair,
the total density of vertices connecting $11$ pairs is
$lm\rho^{{\rm (ss)}}_{11}$.
The densities for other combinations can be obtained in a similar fashion.
Thus, the total density of vertices is
\begin{equation}
  \rho_{lm} \equiv 
  lm\rho^{{\rm (ss)}}_{11} + l\bar m\rho^{{\rm (ss)}}_{10} 
  + \bar l m\rho^{{\rm (ss)}}_{01} + \bar l\bar m\rho^{{\rm (ss)}}_{00}.
  \label{GeneralDensity}
\end{equation}
for two neighboring segments with spin values $l$ and $m$.

Next, we consider the scattering probability of worms at a vertex 
with no kink on it.
Suppose a spin-lowering worm hits the lower-left leg of the vertex from below
in the coarse-grained picture.
\deffig{Scattering}{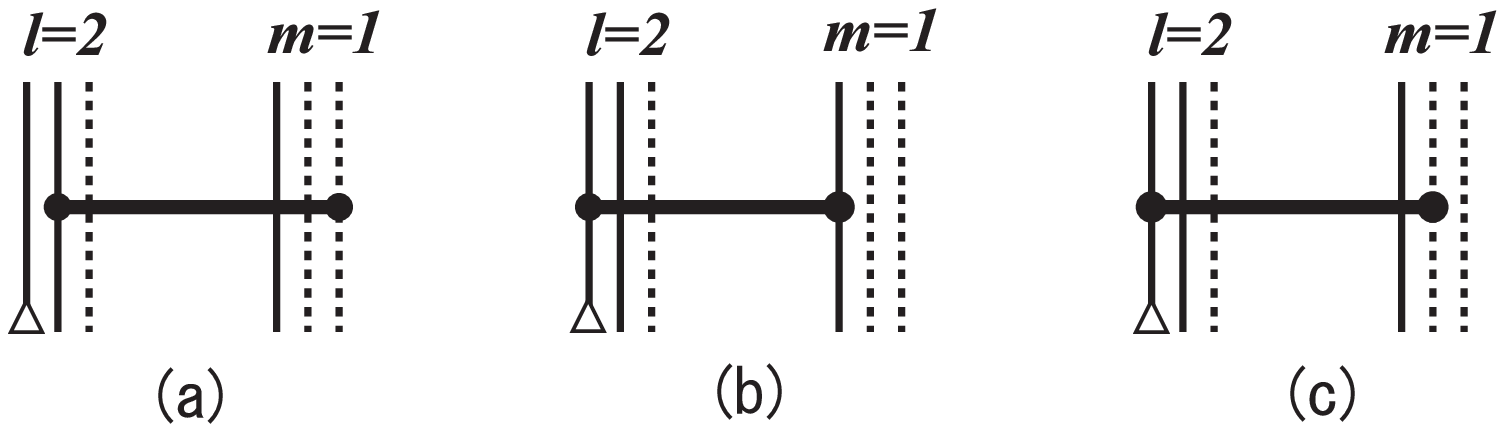}{0.45}{
Scattering at a vertex for $S=3/2$ in the split-spin representation.
The thick horizontal lines represent vertices.
The spin-lowering worm is indicated by an open triangle.
A solid line represents a world line with $n_{i\mu}(\tau)=1$
whereas a dotted one represents $n_{i\mu}(\tau)=0$.
In case (a), the worm cannot be scattered.
There are two cases [(b) and (c)] where the worm can be scattered.
}
In order for this worm to be scattered, 
the worm and one of the legs of the vertex 
must be mapped onto the same line by the
inverse map (\Fig{Scattering}).
There are two such cases: the case where the spin value is 1
[\Fig{Scattering}(b)]
on the legs on the other line and the case where it is 0
[\Fig{Scattering}(c)].
In the first case there are $m$ different choices
of the line, whereas we have $\bar m$ choices 
in the second case.
Each individual choice in the first case has 
the weight $\rho^{{\rm (ss)}}_{11}$ whereas
that in the second case has the weight $\rho^{{\rm (ss)}}_{10}$.
Therefore, the probability for choosing the first case is
$m\rho_{11}^{{\rm (ss)}} / \rho_{lm}$,
whereas that for the second case is
${\bar m}\rho_{10}^{{\rm (ss)}} / \rho_{lm}$.
If we choose the first case, the probability with which
the worm is scattered in the direction $\Gamma$ is
$$
  P^{{\rm (ss)}}
  \left(\Gamma\left|
  {\tiny
  \begin{array}{ll} 1 & 1 \\ 1^- & 1 \end{array}
  }
  \right.\right).
$$
The worm scattering probability for the second case is given similarly.
All in all, the probability of a spin-lowering worm being scattered into 
the direction specified by the directed graph $\Gamma$ ($\ne \uparrow$)
becomes
\begin{eqnarray}
  & & 
  P\left(\Gamma \left| \begin{array}{ll} l & m \\ l^- & m \end{array} \right. \right)
  \equiv \nonumber \\
  &  &
  \frac{
  m \rho^{{\rm (ss)}}_{11} 
  P^{{\rm (ss)}}
  \left(\Gamma\left|
  {\tiny
  \begin{array}{ll} 1 & 1 \\ 1^- & 1 \end{array}
  }
  \right.\right) 
  +  \bar m \rho^{{\rm (ss)}}_{10} 
  P^{{\rm (ss)}}
  \left(\Gamma\left|
  {\tiny
  \begin{array}{ll} 1 & 0 \\ 1^- & 0 \end{array} 
  }
  \right.\right)
  }{
  {lm \rho^{{\rm (ss)}}_{11} + l\bar m \rho^{{\rm (ss)}}_{10} 
   + \bar l m \rho^{{\rm (ss)}}_{01} + \bar l \bar m \rho^{{\rm (ss)}}_{00}}
  }.
  \label{GeneralScatteringProbability}
\end{eqnarray}
The probability for going through ($\Gamma = \uparrow$) is
simply equal to $1-$
(probabilities of the three proper scatterings).
The symbol
$$
  \left(\begin{array}{ll} l' & m' \\ l^{\pm} & m \end{array}\right)
$$
denotes the state where the spin states on the upper-left,
upper-right, lower-left, and lower-right legs are
$l', m', l$, and $m$, respectively, and there is an incoming 
spin-raising ($+$) or spin-lowering ($-$) worm on the lower-left leg.
$P^{{\rm (ss)}}(\Gamma|\Sigma)$ is the probability of the worm being 
scattered into the direction $\Gamma$ when the initial state of the 
vertex is $\Sigma$ in the split-spin representation.
This probability coincides with that in the $S=1/2$ case with the
replacement $H_p\rightarrow h$.
The scattering probability of a spin-raising 
worm can be obtained in the same fashion.

The scattering probability at a vertex with a kink is simpler than
that for a vertex with no kink on it, 
because in this case there is at most one type of vertex that 
may lead to a proper scattering (diagonal, horizontal, or turning back).
For example, suppose a particle jumps from left to right at the kink 
at the imaginary time $\tau$, 
and the spin-lowering worm is approaching the vertex on its lower-left leg.
The local state $\Sigma$ is given by
$$
  \Sigma = \left(\begin{array}{ll} l-1 & m+1 \\ l^{-} & m \end{array}\right).
$$
Then, the vertex's lower-left leg must be footed on positive segments 
[$\sigma_{i\mu}(\tau) = 1$] because otherwise no particle can hop to
the neighboring site there.
Similarly, the lower-right, upper-left, and upper-right legs must be footed on
negative, negative, and positive segments, respectively.
There are $l\bar m$ such choices of segments, and all the choices are equally 
probable.
Among them, there are $\bar m$ choices where the lower-left leg is footed on
the segment where the worm is located.
Therefore, the probability of the worm being located on one of the legs of 
the vertex is $m/(l\bar m) = l^{-1}$.
Then, the scattering probability for $\Gamma$'s corresponding to proper
scatterings [i.e., $\Gamma=\uparrow, \nearrow, \downarrow$] is
\begin{equation}
  P
  \left(\Gamma\left|\begin{array}{ll} l-1 & m+1 \\ l^- & m \end{array}\right.\right)
  = 
  l^{-1}P^{({\rm ss})}
  \left(\Gamma\left|\begin{array}{ll} 0 & 1 \\ 1^- & 0 \end{array}\right.\right),
  \label{GeneralScatteringProbabilityII}
\end{equation}
for spin-lowering worms. 
Probabilities for spin-raising worms can be obtained similarly.

Thus, we have described the way we derive
the density of vertices and
the scattering probability of worms from an
algorithm in the split-spin representation.
Although our description above may seem to give an actual 
procedure for coarse-graining mapping and its inverse,
we do not perform these mappings in real simulation.
They are only for deriving the density 
\Eq{GeneralDensity} and the probabilities
\Eq{GeneralScatteringProbability} and
\Eq{GeneralScatteringProbabilityII}.
In the actual simulation, we manipulate only 
coarse-grained variables.

In order to complete the description of the algorithm,
we have to specify the procedure 
for the pair creation and annihilation of worms.
Again, this can be done by the coarse-graining map and its inverse.
In the split-spin representation, the pair creation of worms
is done simply by choosing a point $((i,\mu),\tau)$ in the system
with a uniform probability distribution.
If there is a hole at the chosen point [i.e., $n_{i\mu}(\tau)=0$],
we create a pair of spin-raising worms there.
If there is a particle instead, we create spin-lowering worms.
When coarse grained, this procedure is mapped to choosing
a point from the whole space-time with uniform probability distribution
and creating a pair of spin-raising worms with probability ${\bar l}/(2S)$
or spin-lowering ones with probability $l/(2S)$, where $l$ is
the spin state at the chosen point.

The moving worm travels according to the scattering process described 
above until it comes back to the original position $((i,\mu),\tau)$
where the other worm waits.
When coarse grained, this ``coming-home'' event is mapped to an event
in which a worm comes back to $(i,\tau)$.
However, several other split-spin events are mapped 
to this same coarse-grained event.
Namely, there are cases where the moving worm comes to 
the point corresponding to a different $\sigma$ spin, 
i.e., $((i,\nu),\tau)$ with $\nu\ne\mu$.
Worms in this case should not annihilate.
It has to be mapped, therefore, to a ``going-through'' event.
Suppose that the worms are spin-lowering ones and that
the local value of the coarse-grained spin is $l$ (before the passage
of the worm).
Then, there are $l$ cases in total which are mapped to 
the same coarse-grained state.
Only one of them leads to the collision of two worms.
Therefore, the probability of pair annihilation is $l^{-1}$
and that for going through is $1-l^{-1}$.
For the same reason,
the probability of annihilation should be $({\bar l})^{-1}$
if the worms are spin-raising ones.

The whole procedure of one Monte Carlo sweep (MCS)
with the algorithm described in this section 
can be summarized as follows.
\begin{description}
\item[Step 1:]
  Place vertices at random with the density, $\rho(\Sigma)$,
  that depends on the local spin state, $\Sigma$.
  Set $n_{{\rm count}} = 0$.
\item[Step 2:]
  Increase $n_{{\rm count}}$ by 1.
  Choose a point in the whole space-time at random, 
  and create two worms there, one is to move and
  the other is to stay.
  For the moving worm,
  choose the initial direction of motion, upward or downward, 
  with the probability 1/2.
  Then choose its initial type, spin-lowering or spin-raising,
  with the probability $l/(2S)$ or $\bar l/(2S)$, respectively.
\item[Step 3:]
  Let the moving worm go until it hits a vertex or comes
  back to the original position where the other worm stays.
  If it hits a vertex before it comes back to the original position,
  go to step 4.
  Otherwise, go to step 5.
\item[Step 4:]
  Choose the scattering direction $\Gamma$ with the probability
  $P(\Gamma|\Sigma)$, where $\Sigma$ is the local spin state at
  the vertex before the worm's arrival.
  Change the type of the worm as specified by $\Gamma$.
  Then, go back to step 3.
\item[Step 5:]
  If the moving worm is a spin-lowering one,
  let it go through the original point 
  with the probability $1-l^{-1}$ and go to step 3.
  Otherwise let it go through
  with the probability $1- {\bar l}^{-1}$ and go to step 3.
  If the moving worm does not go through, 
  let it annihilate with its partner and go to step 6.
\item[Step 6:]
  If $n_{{\rm count}}$ is smaller than $n_{{\rm max}}$, go back to step 2.
  Otherwise, erase all the vertices with no kink and go back to step 1.
\end{description}
One Monte Carlo sweep is defined as a process between two successive
resets of vertices (i.e., two successive passages of step 1).
The number $n_{\rm max}$,
the number of pair creations of worms during 1 MCS,
can be an arbitrary positive integer.
We choose it so that every vertex may be visited by a worm once in
average during 1 MCS.

\section{The $XXZ$ models}
\label{sec:XXZ}
Since the $XXZ$ models are of particular importance,
we summarize the probability of vertices and
the scattering probability of worms for the models
in \Tab{ScatteringProbability}.
Besides the coupling constants,
the scattering probability depends upon the initial configuration
of the scatterer (i.e., vertex), 
the type of the worm (``spin-raising'' or ``spin-lowering''),
and the incoming and outgoing direction.
Because of the mirror image symmetries with respect to the 
horizontal and vertical axes, 
scattering probabilities for any two cases which can be 
transformed to each other by mirror image transformations
should be the same.
Therefore, without loss of generality, we assume that the incoming 
worm is located on the lower-left leg of the vertex.
Then, the initial states
can be categorized into six classes, 
each specified by the spin states on all the legs 
and the kind of the incoming worm
(\Tab{ScatteringProbability}).

There are only a few possible final states for each initial state.
Those final states can be specified by the outgoing direction 
($\Gamma$) of the scattered worm (\Fig{FinalStates}).
\deffig{FinalStates}{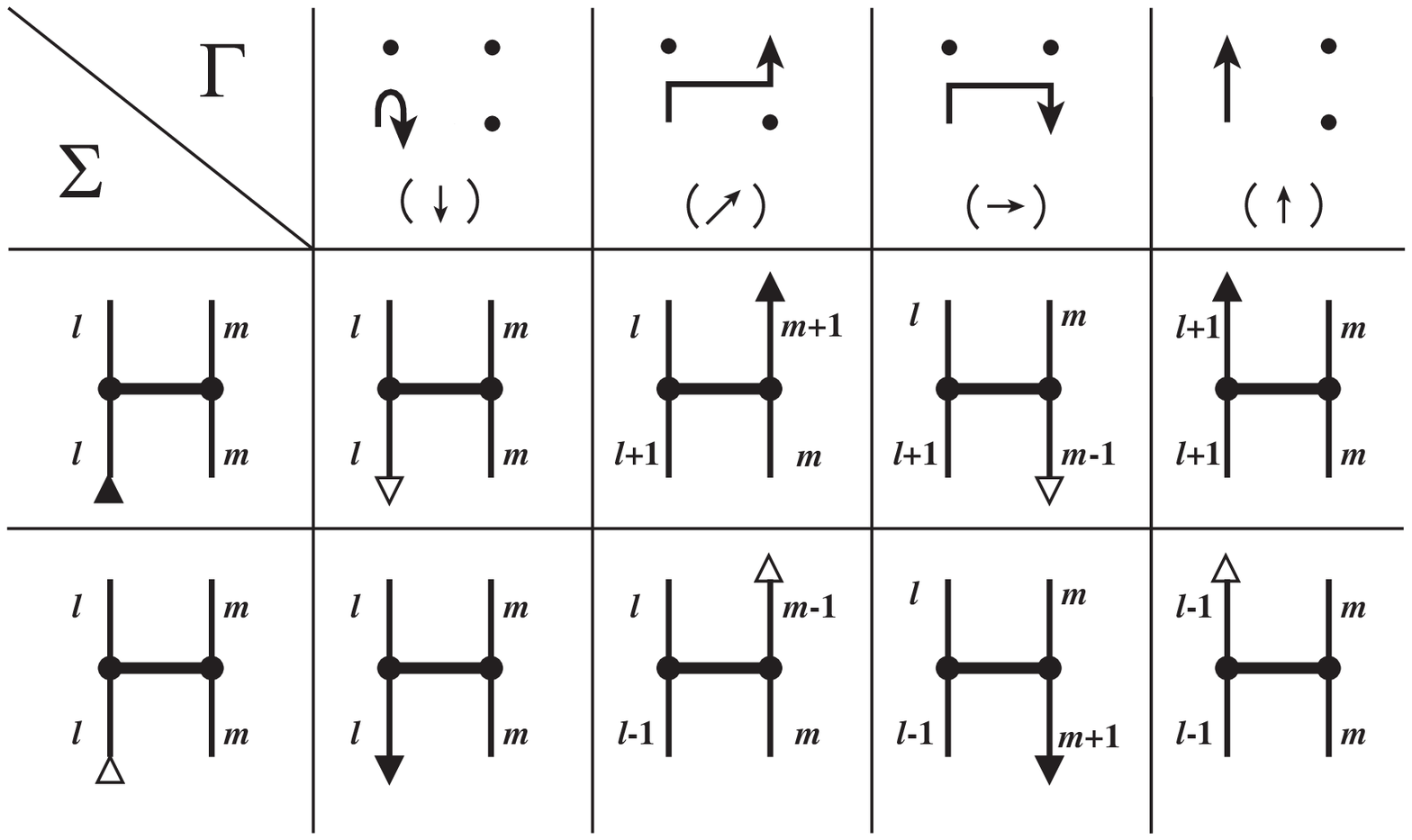}{0.5}{
Some examples of the final state $\Sigma^{\Gamma}$ of scattering
for which the initial state is $\Sigma$ and the outgoing direction
of the worm is specified by the directed graph $\Gamma$.
A solid triangle denotes a spin-raising worm whereas a open triangle a
spin-lowering one.
The dots in the directed graphs represent segments for which
spin variables are not changed by the scattering.
The symbols in the parentheses are abbreviated forms of $\Gamma$.
}
There are four such directions:
turning-back, diagonal, horizontal, and straight,
as indicated in the top row in \Fig{FinalStates}.
The probabilities for scattering in these directions are denoted by
$ P(\downarrow|\Sigma),
  P(\nearrow|\Sigma),
  P(\rightarrow|\Sigma),
$
and
$
  P(\uparrow|\Sigma)
$
respectively, where $\Sigma$ is the local state in
the coarse-grained spin representation.
In \Tab{ScatteringProbability}, we present the first three only.
The probability for going straight $P(\uparrow|\Sigma)$ 
can be readily obtained by subtracting 
the other three from unity.

The scattering probability also depends
upon the coupling constants, $J$ and $J'$.
From the algorithmic point of view,
the whole parameter space is divided into six regions (\Fig{PhaseDiagram}).
\deffig{PhaseDiagram}{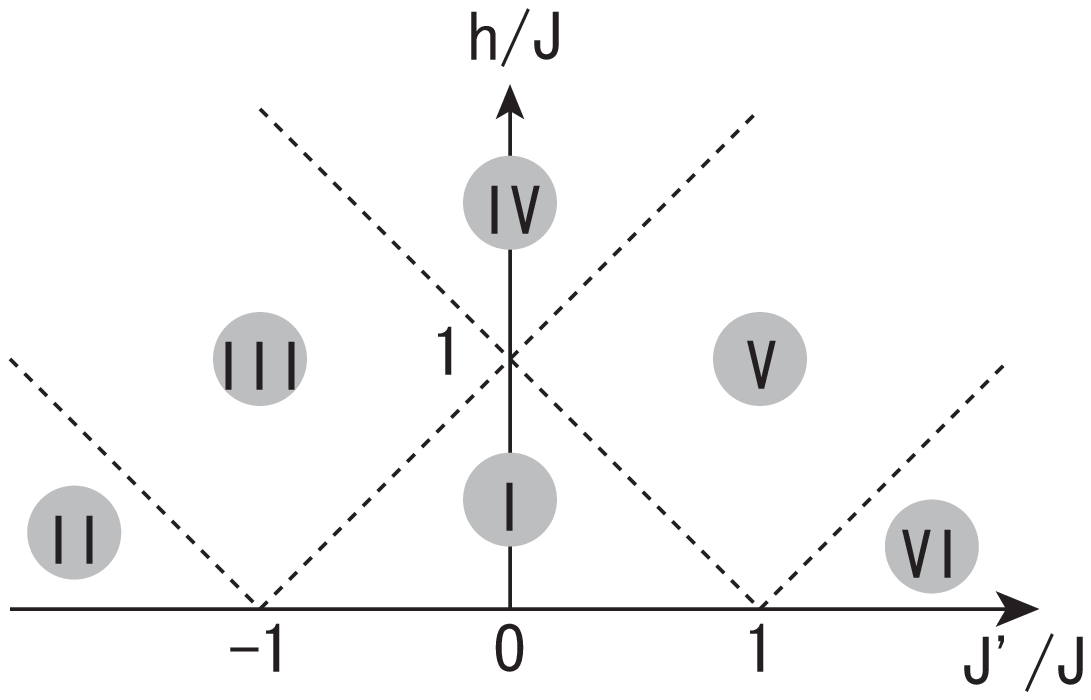}{0.35}{
The six regions in the parameter space for the $XXZ$ model
with general $S$.
The same as the one in Sylju\r{a}sen and Sandvik's paper
\cite{SyljuasenS2002} for $S=1/2$.
}
Within each one of the six regions, the scattering probability
is a simple analytic function of the coupling constant,
and it is continuous in the whole parameter space.
However, its first derivatives are discontinuous 
at the boundary between two adjacent regions.
In the case of $S=1/2$, the division is
the same as that in the previous paper\cite{SyljuasenS2002}.

\begin{table*}
\caption{
The coarse-grained algorithm for the $XXZ$ spin models.
The density of vertices $\rho$, and 
the scattering probabilities of worms $P$.
$h \equiv H_p/(2S)$, $\bar l \equiv 2S - l$, and $\bar m \equiv 2S - m$.
}
\label{tab:ScatteringProbability}
\begin{ruledtabular}
\begin{tabular}{c|lcccccc}
$\Sigma$ &  & Region I & Region II & Region III & Region IV & Region V & Region VI \\
\hline
\hline
$\left( \begin{array}{cccc} l & m \\ l & m \end{array} \right)$ &
$\rho(\Sigma)= $ &
A & B & B & B & A & C \\
\hline
\hline
  &
$P(\downarrow|\Sigma)=$ &
0 & 
$\frac{m (-J-J'-h)}{2B}$ &
0 &
0 &
0 &
$\frac{\bar m (-J+J'-h)}{2C}$ \\
$\left(\begin{array}{cccc} l & m \\ l^+ & m \end{array}\right)$ &
$P(\nearrow|\Sigma)=$ &
$\frac{\bar m (J+J'-h)}{4A}$ & 
0 &
0 &
0 &
$\frac{\bar m (J+J'-h)}{4A}$ &
$\frac{\bar m J}{2C}$ \\
  &
$P(\rightarrow|\Sigma)=$ &
$\frac{m (J-J'-h)}{4A}$ &
$\frac{m J}{2B}$ &
$\frac{m (J-J'-h)}{4B}$ &
0 &
0 &
0 \\
\hline
  &
$P(\downarrow|\Sigma)=$ &
0 &
$\frac{\bar m (-J-J'+h)}{2B}$ &
$\frac{\bar m (-J-J'+h)}{2B}$ &
$\frac{
{\scriptsize
\begin{array}{ll}
m (-J+J'+h) \\
\quad + \bar m (-J-J'+h)
\end{array}
}
}{2B}$
&
$\frac{m (-J+J'+h)}{2A}$ &
$\frac{m (-J+J'+h)}{2C}$ \\
$\left(\begin{array}{cccc} l & m \\ l^- & m \end{array}\right)$ &
$P(\nearrow|\Sigma)=$ &
$\frac{m (J+J'+h)}{4A}$ &
0 &
$\frac{m (J+J'+h)}{4B}$ &
$\frac{m J}{2B}$ &
$\frac{m J}{2A}$ &
$\frac{m J}{2C}$ \\
  &
$P(\rightarrow|\Sigma)=$ &
$\frac{\bar m (J-J'+h)}{4A}$ &
$\frac{\bar m J}{2B}$ &
$\frac{\bar m J}{2B}$ &
$\frac{\bar m J}{2B}$ &
$\frac{\bar m (J-J'+h)}{4A}$ &
0 \\
\hline
  &
$P(\downarrow|\Sigma)=$ &
0 &
0 &
0 &
0 &
0 &
0 \\
$\left(\begin{array}{cccc} l+1 & m \\ l^+ & m+1 \end{array}\right)$ &
$P(\nearrow|\Sigma)=$ &
$\frac{J+J'+h}{\bar l\cdot 2J}$ &
0 &
$\frac{J+J'+h}{\bar l\cdot 2J}$ &
$\frac1{\bar l}$ &
$\frac1{\bar l}$ &
$\frac1{\bar l}$ \\
  &
$P(\rightarrow|\Sigma)=$ &
$\frac{J-J'-h}{\bar l\cdot 2J}$ &
$\frac1{\bar l}$ &
$\frac{J-J'-h}{\bar l\cdot 2J}$ &
0 &
0 &
0 \\
\hline
  &
$P(\downarrow|\Sigma)=$ &
0 &
0 &
0 &
0 &
0 &
0 \\
$\left(\begin{array}{cccc} l-1 & m \\ l^- & m-1 \end{array}\right)$ &
$P(\nearrow|\Sigma)=$ &
$\frac{J+J'-h}{l\cdot 2J}$ &
0 &
0 &
0 &
$\frac{J+J'-h}{l\cdot 2J}$ &
$\frac1l$ \\
  &
$P(\rightarrow|\Sigma)=$ &
$\frac{J-J'+h}{l\cdot 2J}$ &
$\frac1l$ &
$\frac1l$ &
$\frac1l$ &
$\frac{J-J'+h}{l\cdot 2J}$ &
0 \\
\hline
$\left(\begin{array}{cc} l+1 & m \\ l^- & m+1 \end{array} \right)$ &
  &
\multicolumn{6}{l}{
$
P(\downarrow|\Sigma)=
P(\nearrow|\Sigma)=
P(\rightarrow|\Sigma)= 0, \quad\mbox{and}\quad
P(\uparrow|\Sigma)=1
$
}
\\
\hline
$\left(\begin{array}{cc} l-1 & m \\ l^+ & m-1 \end{array} \right)$ &
  &
\multicolumn{6}{l}{
$
P(\downarrow|\Sigma)=
P(\nearrow|\Sigma)=
P(\rightarrow|\Sigma)= 0, \quad\mbox{and}\quad
P(\uparrow|\Sigma)=1
$
}
\\
\hline
\hline
\multicolumn{8}{c}{
\begin{minipage}{130mm}
\begin{eqnarray*}
  A & \equiv & 
    \frac14 [lm (J+J'+3h) 
    + (l\bar m + \bar l m)(J-J'+h) + \bar l \bar m (J+J'-h)] \\
  B & \equiv &
    lm h + (l\bar m + \bar l m) \frac{-J'+h}2, \qquad
  C \equiv  \frac12 [lm(J'+h) + \bar l \bar m (J'-h)] 
\end{eqnarray*}
\end{minipage}
}
\end{tabular}
\end{ruledtabular}
\end{table*}

\section{Efficiency}
\label{sec:Performance}
It is practically impossible to evaluate the efficiency of the algorithm
for all possible combinations of coupling constants, the external field,
the magnitude of spins, and the number of dimensions.
Therefore, here we only show an example and
make a few remarks concerning the efficiency
of the algorithm described above.

Of particular interest is the algorithm in region III,
because the primary motivation for developing the algorithms
based on the SS scheme is to solves the freezing problem
of the conventional loop algorithms in this region.
For $S=1/2$, good performance was demonstrated\cite{SyljuasenS2002}
in the isotropic case $|J'| = J$ for various values of $H$.
Most importantly, no severe freezing was observed at low temperature.

In what follows, we show that the present algorithm solve
this problem for an arbitrary $S$.
Several other directed loop algorithms 
(algorithms 1 -- 4) in the original spin representation 
are also examined for comparison with the present algorithm.
Algorithms 1 -- 3 are obtained by tuning solutions of the 
weight equation\Eq{WeightEquation} 
and the detailed balance\Eq{DetailedBalance}
so that the turning-back probabilities may be minimized.
All of these three algorithms have exactly the same turning-back
probabilities.
Algorithm 1 is characterized by the vanishing probability
for the diagonal scattering when the field is zero
[i.e., $\lim_{h\to 0}P(\nearrow|\Sigma)=0$ for all $\Sigma$], 
whereas it is finite even at $h=0$ in algorithm 2.
Algorithm 3 is a mixture of algorithms 1 and 2.
The details of these algorithms are given in \App{Examples}.
Algorithm 4, on the other hand, 
is the heat-bath-type algorithm that can be obtained
in the most straightforward way,
although this is also a solution of Eqs. \Eq{WeightEquation} 
and \Eq{DetailedBalance}.
In algorithm 4, one of the four possible directions $\Gamma$ is 
chosen with the probability proportional to the weight of the final state.
To be specific,
$$
  P(\Gamma|\Sigma) = 
  \frac{W(\Sigma^{\Gamma})}{\sum_{\Gamma'}W(\Sigma^{\Gamma'})},
$$
where $\Sigma$ is the initial state of the vertex and 
$\Sigma^{\Gamma}$ is the final state of the vertex when 
the worm is scattered into the direction $\Gamma$.

In order to check the validity of these algorithms,
we first performed simulations for a small one-dimensional
system ($L=4$) and compared the results with the exact solution
for various set of parameters, $J'$, $H$, and $\beta$.
It turned out that all the algorithms yielded correct results
with 1\% or less of the statistical error.
The present algorithm and algorithm 1 yielded roughly the
same magnitude of error whereas the other three yielded larger 
errors than the first two.

For a longer chain ($L=64$), 50 sets of simulations were performed 
using each algorithm where each set consists of 20000 creations and 
annihilations of worm pairs.
We can see the performances of five algorithms in \Fig{StatisticalError}.
Plotted in \Fig{StatisticalError} is $\Delta (M_{\pi}^2)N_v^{1/2}/L$,
where $\Delta (M_{\pi}^2)$ is the estimated statistical error 
of the squared staggered magnetization, $L$ is the system size 
and $N_v$ is the average number of the vertices visited by the worm
during its lifetime.
Since the scattering process is the most time-consuming part of the code, 
the total CPU time is roughly proportional to the total number of
scattering events of worms, including the ``straight'' scatterings.
Therefore, the CPU time is proportional to $N_v$.
This is why the statistical error should be multiplied by $N_v^{1/2}$
in order to make the comparison fair.
In \Fig{StatisticalError}, we can clearly see that the
present algorithm performs as well as the best algorithm among the
the other four (i.e., algorithm 1).
Obviously, there is no exponential slowing down for the present algorithm
and algorithm 1, as was the case with Sylju\r{a}sen and Sandvik's
algorithm for $S=1/2$.
\deffig{StatisticalError}{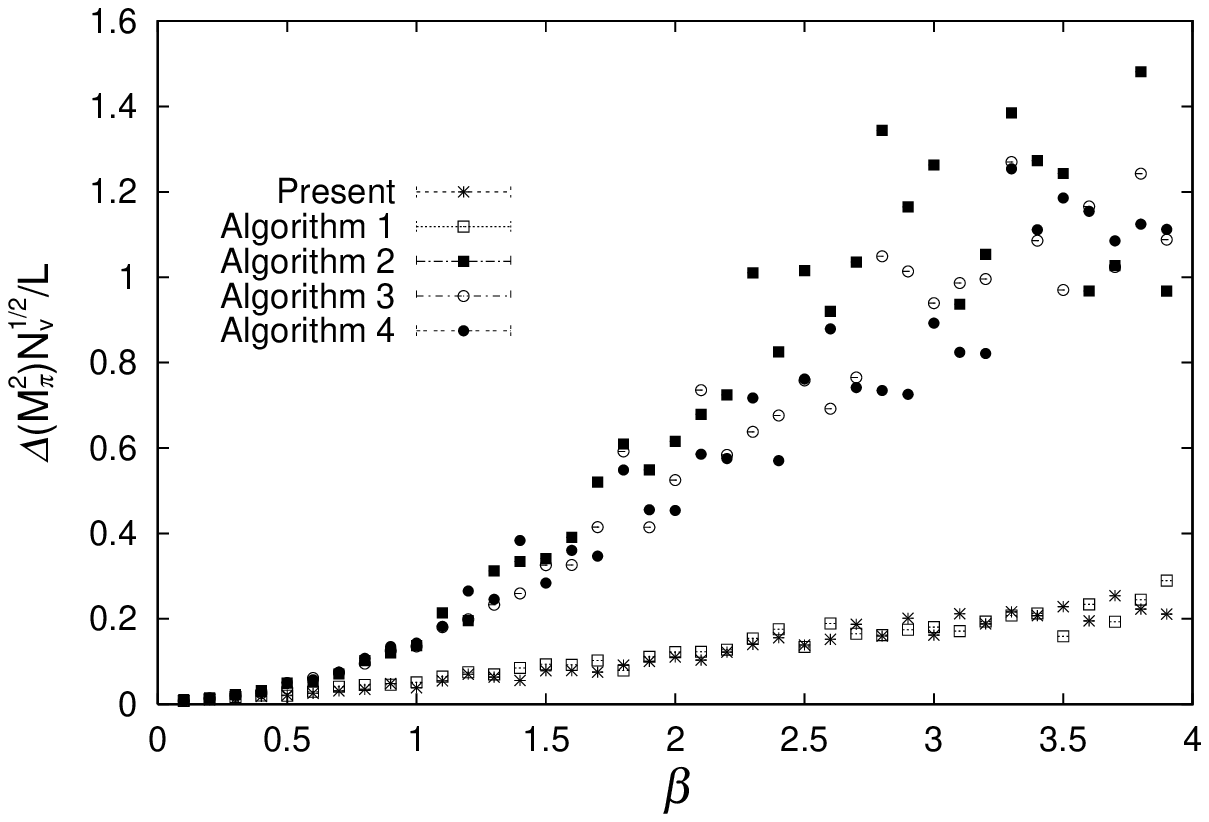}{0.5}{
The statistical error in the estimate of the squared staggered 
magnetization multiplied by the square root of the average number of 
scattering events during the lifetime of a worm.
The system is the $S=1$ antiferromagnetic
Heisenberg chain of length $L=64$ 
with a uniform magnetic field $H=0.1$.
Each point is a result of 50 sets of simulations, where each set 
consists of 20000 pair creations and annihilations of worms.
}

For a larger system ($L=64$) with zero magnetic field, 
the results of the present algorithm and algorithm 1
agree with each other, while those of the other algorithms do not
(\Fig{SystematicError}).
We consider that the correct result for $L=64$ is the one that 
is obtained by the present algorithm and algorithm 1, and that
the other algorithms fail to achieve equilibrium within 
the performed simulation. 
\deffig{SystematicError}{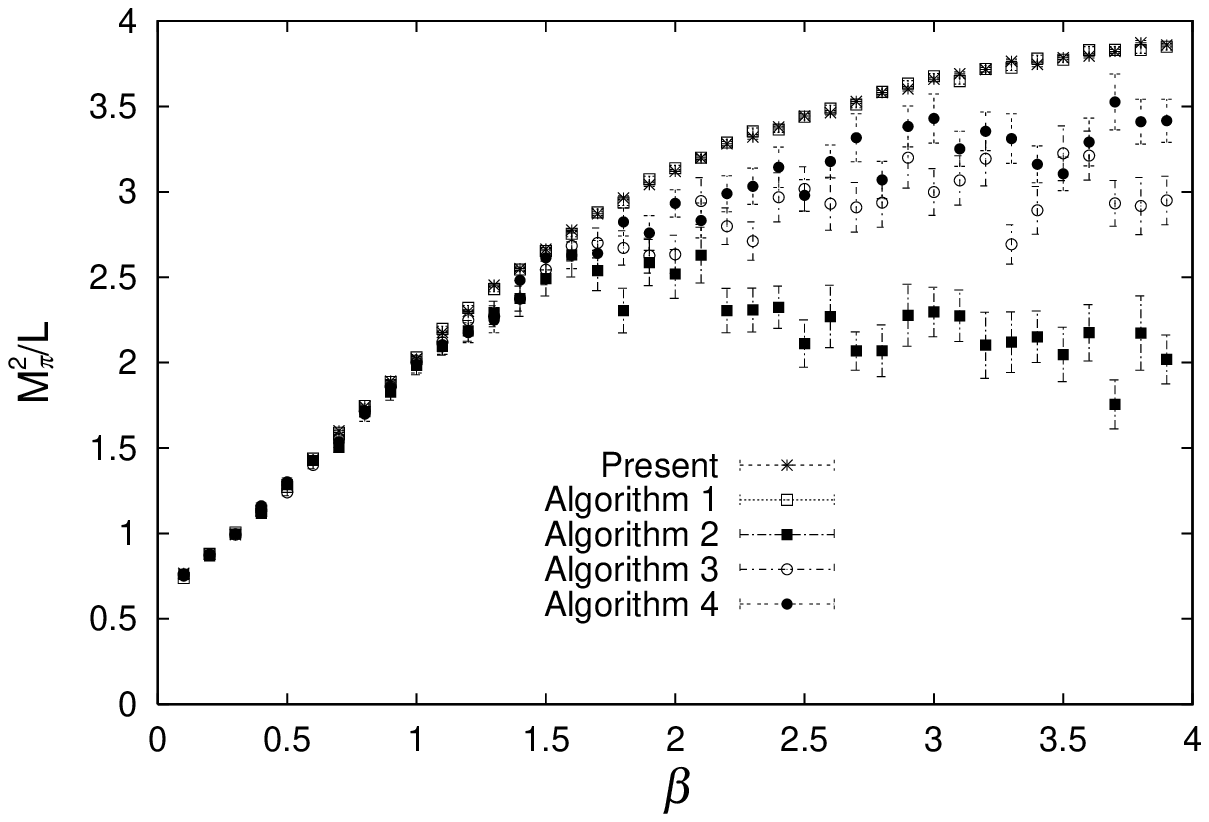}{0.5}{
The squared staggered magnetization estimated
with the present algorithm and those with the other algorithms
for the $S=1$ antiferromagnetic Heisenberg chain with $L=64$ and $H=0$.
The rest of the condition for the simulation is the same as that
for the previous figure.
}
We performed similar simulations for various values of $H$
ranging from $H=0.0$ to $H=2.0$.
It turned out that the good algorithms (the present and algorithm 1) 
always perform better than the bad ones (algorithm 2 -- 4)
although the difference between them becomes smaller 
as the field is increased.

We can explain these facts in terms of the compatibility of
the clusters with the order parameter.
One of the reasons why the conventional loop algorithm works well
even in the vicinity of the critical point is the accordance between
the typical cluster size and the correlation length.
This accordance is guaranteed by the two features of the algorithm:
(i) independent flipping of clusters, and
(ii) perfect ordering within each cluster.
Although the staggered magnetization is not strictly the order parameter
in one dimension, this criterion of good performance of loop-cluster
algorithms still applies because finite but relatively long-range
correlation exists even in one dimension.
It is easy to see that the present algorithm and algorithm 1 satisfy
both conditions (i) and (ii) in the zero-field limit whereas 
the other algorithms do not satisfy condition (ii) regardless of the field.
This is the reason why the former two algorithms perform much better
than the latter three in the weak field region.
Therefore, we expect that the difference in the efficiency
is even more pronounced near real phase transitions such as these
in three-dimensional systems.

The results of the five algorithms, all based on the SS scheme,
illustrate that it is not trivial in general to obtain 
the most efficient algorithm among many possible ones and also that
the straightforward heat-bath algorithm is rather poor 
in some important cases.
It should be noted that the coarse-grained algorithm
discussed in the present paper satisfies criterion (ii)
mentioned above for an arbitrary $XXZ$ model when the external
field is vanishing.
Therefore, we consider that the present algorithm performs well
for a rather wide class of models in the weak magnetic field
region.
Even for strong magnetic field, we consider that the present 
algorithm is at least as good as most of the other algorithms 
based on the SS scheme as we see above.

For the performance of the present algorithm in the regions other than III, 
we cannot conclude much at the moment.
In the case of $S=1/2$,
the algorithm in region I is equivalent to the Wolff version 
(i.e., the single-loop update version) of the loop algorithm
when the external field is vanishing.
A good performance of the algorithm in this region has been demonstrated
in many applications\cite{WieseY},
as well as for $S>1/2$ \cite{KawashimaG1995,HaradaTKXY}.
Since the turning-back probability is vanishing in the whole region I,
we expect good performance of the algorithm not only on the line $h=0$
but also in the whole region I.
The algorithm in regions II and VI, on the other hand,
may not be very efficient because of the presence of the relatively 
large turning-back probability.
It is easy to see that, in the classical limit ($J'/J \to \infty$)
at zero external field, 
the turning-back probability dominates in regions II and VI,
leading to poor performance.
For region VI, this may not be very problematic because in this region
(at least in the classical Ising limit)
the conventional loop-cluster algorithm works efficiently.
For region II, it is not known whether this is a real problem or not.
The algorithm in region IV is reduced to a single-spin-flip 
Metropolis algorithm in the limit of $J'/J \to 0$ and $h \to \infty$.
The performance of the Metropolis algorithm should be good in this limit,
although the region is physically not very interesting.

\section{Summary}
We have proposed an algorithms based on the Sylju\r{a}sen and Sandvik scheme
by introducing the split-spin representation and the coarse-graining procedure.
The algorithm is a natural extension of the directed loop algorithm for
$S=1/2$, in that the present algorithm coincides with it for $S=1/2$.
In addition, the present algorithm is a natural extension of
the conventional loop algorithm,
because if the external field is vanishing,
the present algorithm can be obtained through coarse graining
the conventional split-spin loop algorithm.

Compared with the algorithms in the split-spin representation,
the coarse-grained algorithm requires much smaller memory, in general.
In particular, when the algorithm consists of
vertices with more than four legs, as is generally the case with 
the loop algorithms for models with high order interaction terms,
the memory can be reduced considerably.
The coarse-grained algorithm is also advantageous since the
codes based on it can be very easily modified for other models
(we only need to change the arrays of the probability tables).

In the case of the $S=1$ Heisenberg antiferromagnet in one dimension,
the algorithm's performance is almost the same as the best
algorithm obtained by directly working on the 
original spin representation.
Many other algorithms can also be obtained in the same way.
However, most of them, including the heat-bath algorithm,
are much worse than the present one.
Existence of algorithm 1, a good direct solution
to Eqs. \Eq{WeightEquation} and \Eq{DetailedBalance},
suggests the existence of similar solutions for an arbitrary $S$.
We have not succeeded in finding a complete set of such solutions,
although we believe that such solutions exist.
It would be an interesting future problem to find such solutions
for various models.

The coarse-graining procedure presented in this paper
applies not only to the $XXZ$ spin systems but also
to any model for which a directed loop algorithm can be constructed
in the split-spin representation.
For example, on-site easy axis or easy plane anisotropy terms may be treated
as the couplings between $\sigma$ spins on the same site.
Another example is the SU(N) model where the split-spin algorithm is known.
 

\begin{acknowledgments}
The authors are grateful to O.~Sylju\r{a}sen,
A.~W.~Sandvik, and M.~Troyer for 
enlightening comments.
The present work is supported by Grants-in-Aid for Scientific 
Research Program (No. 12740232 and No. 14540361) from Monkasho, Japan.
\end{acknowledgments}

\appendix
\section{General formulation of the SS scheme}
\label{sec:Framework}

In general, a directed loop algorithm is characterized by the density of
vertices $\rho(\Sigma)$ and the scattering probability of worms 
$P(\Gamma|\Sigma)$.
The density $\rho(l,m)$ is simply given by
\begin{equation}
  \rho(l,m) = W\left(\begin{array}{ll} l & m \\ l & m \end{array}\right),
  \label{Density}
\end{equation}
where the weight $W(\Sigma)$ is defined as
\begin{eqnarray}
  W\left(\begin{array}{ll} l' & m' \\ l^{\epsilon} & m \end{array}\right) 
  & \equiv &
  \left(
  c\delta_{ll'}\delta_{mm'} - \langle l',m' | H_{ij} | l,m \rangle
  \right) \nonumber \\
  & & \qquad \times\Delta(0\le l+\epsilon \le 2S),
  \label{Weight}
\end{eqnarray}
where $\Delta(``\cdots") = 1$ when $``\cdots"$ is true and
$0$ otherwise.
The symbol $\epsilon$ stands for the integer by which
the worm changes the spin value, e.g., $\epsilon = -2$ for
a $(--)$ worm that lowers the spin value by 2.
The variable $c$ is the only free parameter related to 
the vertex density.

The scattering probability, on the other hand, has a lot of freedom.
The algorithm can be explained very clearly by introducing an extended
weight $W(\Sigma,\Gamma)$ that is related to the weight $W(\Sigma)$ as
\begin{equation}
  W(\Sigma) = \sum_{\Gamma} W(\Sigma,\Gamma).
  \label{WeightEquation}
\end{equation}
Here, we consider a scattering event in which the initial state of
the vertex is $\Sigma$ and the out going direction of the worm is 
$\Gamma$.
We denote the final state of this event as $\Sigma^{\Gamma}$.
It should be noted that the `state' $\Sigma$ is directed
in contrast to the state in the ordinary Monte Carlo simulation.
We consider the balance between an arbitrary sequence of 
scatterings and its reverse.
Each sequence starts from pair creation of the worms
and ends at pair annihilation,
The detailed balance condition should be considered 
between $\Sigma$ and the reverse of $\Sigma^{\Gamma}$.
Therefore, the detailed balance condition is expressed as
\begin{equation}
  W(\Sigma,\Gamma) = W(\bar{\Sigma^{\Gamma}},\bar{\Gamma}).
  \label{DetailedBalance}
\end{equation}
Here, $\bar{\Sigma^{\Gamma}}$ is the reverse of $\Sigma^{\Gamma}$ obtained 
by inverting the direction and changing the type of the worm 
in $\Sigma^{\Gamma}$, whereas $\bar{\Gamma}$ is the inverse of $\Gamma$, 
obtained by inverting the direction of the arrow in $\Gamma$.

It is worth mentioning that Eqs. \Eq{WeightEquation} and \Eq{DetailedBalance}
are quite similar to the equations that appear in the general formulations
of the loop-cluster algorithm\cite{KandelD1991,KawashimaG1995}.
The only difference is that here we use directed graphs and directed states
whereas only nondirected graphs and states appear 
in the conventional loop-cluster algorithm.
It is easy to see that in the case of zero magnetic field
the extended weight in the present scheme 
can be made independent of the directions of states and graphs, 
and all the equations discussed in this appendix coincide with 
those for conventional loop-cluster algorithms.

Once we obtain any set of constant $c$ and positive $W(\Sigma,\Gamma)$'s
that satisfy Eqs. \Eq{Weight}, \Eq{WeightEquation}, and \Eq{DetailedBalance},
we obtain a scattering probability $P(\Gamma|\Sigma)$ of worms that satisfies 
the detailed balance condition as
\begin{equation}
  P(\Gamma|\Sigma) = \frac{W(\Sigma,\Gamma)}{W(\Sigma)}.
  \label{ScatteringProbability}
\end{equation}

\section{Some directed loop algorithms for the $S=1$ 
antiferromagnetic Heisenberg model}
\label{sec:Examples}
In Tables \ref{tab:PMOneWeightForAFH} and \ref{tab:PMTwoWeightForAFH}, 
we show the weight and the extended weight
that satisfy  Eqs.
\Eq{Weight}, \Eq{WeightEquation}, and \Eq{DetailedBalance}
for the $S=1$ antiferromagnetic Heisenberg model.
The vertex density and the worm-scattering probability can be obtained
through Eqs. \Eq{Density} and \Eq{ScatteringProbability}.

\Tab{PMOneWeightForAFH} shows the scattering probability for worms that
change values of spin by 1.
The weights contain two free parameters $A = J-B$ and $A' = J-B'$.
Note, however, that the turning-back probability does not depend on 
the choice of the free parameters.
Algorithm 1 -- 3 correspond to the following choices, respectively:
\begin{eqnarray*}
\mbox{algorithm 1,} & & A = \frac{H_p}{4}, \quad A' = J, \\
\mbox{algorithm 2,} & & A = J,\quad A' = \frac{H_p}{4}, \\
\mbox{algorithm 3,} & & A = 0.9J+0.1\frac{H_p}{4}, \quad 
                        A' = 0.9\frac{H_p}{4}+0.1J. \\
\end{eqnarray*}
\begin{table*}
\caption{
  A $\pm 1$ worm solution ($W(\Sigma,\Gamma)$) 
  to the detailed balance equation for the $S=1$
  antiferromagnetic Heisenberg model.
  Applies only if $0\le H_p\le 4J$.
  Free parameters $A, B, A'$, and $B'$ are related to each other by
  $A+B = A'+B' = J$.
}
\label{tab:PMOneWeightForAFH}
\begin{ruledtabular}
\begin{tabular}{cc|cccc||cc|cccc}
  $\Sigma$ & $W(\Sigma)$ & $\downarrow$ & $\uparrow$ & $\nearrow$ & $\rightarrow$ &
  $\Sigma$ & $W(\Sigma)$ & $\downarrow$ & $\uparrow$ & $\nearrow$ & $\rightarrow$ \\
\hline
  $\left(\begin{array}{ll} 2 & 2 \\ 2^+ & 2 \end{array}\right)$ & $0$ &
  0 & 0 & 0 & 0 &
  $\left(\begin{array}{ll} 2 & 2 \\ 2^- & 2 \end{array}\right)$ & $2H_p$ &
  0 & $\frac{7H_p}{4}$ & $\frac{H_p}{4}$ & 0 \\
  $\left(\begin{array}{ll} 2 & 1 \\ 2^+ & 1 \end{array}\right)$ & $0$ &
  0 & 0 & 0 & 0 &
  $\left(\begin{array}{ll} 2 & 1 \\ 2^- & 1 \end{array}\right)$ & $J+\frac{3H_p}{2}$ &
  0 & $\frac{5H_p}{4}$ & $A$ & $B+\frac{H_p}{4}$ \\
  $\left(\begin{array}{ll} 2 & 0 \\ 2^+ & 0 \end{array}\right)$ & $0$ &
  0 & 0 & 0 & 0 &
  $\left(\begin{array}{ll} 2 & 0 \\ 2^- & 0 \end{array}\right)$ & $2J+H_p$ &
  $\frac{H_p}{2}$ & $J+\frac{H_p}{2}$ & 0 & $J$ \\
  $\left(\begin{array}{ll} 1 & 2 \\ 1^+ & 2 \end{array}\right)$ & $J+\frac{3H_p}{2}$ &
  0 & $\frac{7H_p}{4}$ & 0 & $J-\frac{H_p}{4}$ &
  $\left(\begin{array}{ll} 1 & 2 \\ 1^- & 2 \end{array}\right)$ & $J+\frac{3H_p}{2}$ &
  0 & $J+\frac{5H_p}{4}$ & $\frac{H_p}{4}$ & 0 \\
  $\left(\begin{array}{ll} 1 & 1 \\ 1^+ & 1 \end{array}\right)$ & $J+H_p$ &
  0 & $\frac{5H_p}{4}$ & $A-\frac{H_p}{4}$ & $B$ &
  $\left(\begin{array}{ll} 1 & 1 \\ 1^- & 1 \end{array}\right)$ & $J+H_p$ &
  0 & $\frac{3H_p}{4}$ & $B'+\frac{H_p}{4}$ & $A'$ \\
  $\left(\begin{array}{ll} 1 & 0 \\ 1^+ & 0 \end{array}\right)$ & $J+\frac{H_p}{2}$ &
  0 & $J+\frac{H_p}{2}$ & 0 & 0 &
  $\left(\begin{array}{ll} 1 & 0 \\ 1^- & 0 \end{array}\right)$ & $J+\frac{H_p}{2}$ &
  $\frac{H_p}{2}$ & 0 & 0 & $J$ \\
  $\left(\begin{array}{ll} 0 & 2 \\ 0^+ & 2 \end{array}\right)$ & $2J+H_p$ &
  0 & $J+\frac{5H_p}{4}$ & 0 & $J-\frac{H_p}{4}$ &
  $\left(\begin{array}{ll} 0 & 2 \\ 0^- & 2 \end{array}\right)$ & $0$ &
  0 & 0 & 0 & 0 \\
  $\left(\begin{array}{ll} 0 & 1 \\ 0^+ & 1 \end{array}\right)$ & $J+\frac{H_p}{2}$ &
  0 & $\frac{3H_p}{4}$ & $B'$ & $A'-\frac{H_p}{4}$ &
  $\left(\begin{array}{ll} 0 & 1 \\ 0^- & 1 \end{array}\right)$ & $0$ &
  0 & 0 & 0 & 0 \\
  $\left(\begin{array}{ll} 2 & 1 \\ 1^+ & 2 \end{array}\right)$ & $J$ &
  0 & 0 & $\frac{H_p}{4}$ & $J-\frac{H_p}{4}$ &
  $\left(\begin{array}{ll} 2 & 1 \\ 1^- & 2 \end{array}\right)$ & $J$ &
  0 & $J$ & 0 & 0 \\
  $\left(\begin{array}{ll} 0 & 1 \\ 1^+ & 0 \end{array}\right)$ & $J$ &
  0 & $J$ & 0 & 0 &
  $\left(\begin{array}{ll} 0 & 1 \\ 1^- & 0 \end{array}\right)$ & $J$ &
  0 & 0 & 0 & $J$ \\
  $\left(\begin{array}{ll} 1 & 2 \\ 2^+ & 1 \end{array}\right)$ & 0 &
  0 & 0 & 0 & 0 &
  $\left(\begin{array}{ll} 1 & 2 \\ 2^- & 1 \end{array}\right)$ & $J$ &
  0 & 0 & $A-\frac{H_p}{4}$ & $B+\frac{H_p}{4}$ \\
  $\left(\begin{array}{ll} 1 & 0 \\ 0^+ & 1 \end{array}\right)$ & $J$ &
  0 & 0 & $B'+\frac{H_p}{4}$ & $A'-\frac{H_p}{4}$ &
  $\left(\begin{array}{ll} 1 & 0 \\ 0^- & 1 \end{array}\right)$ & $0$ &
  0 & 0 & 0 & 0 \\
  $\left(\begin{array}{ll} 2 & 0 \\ 1^+ & 1 \end{array}\right)$ & $J$ &
  0 & 0 & $\frac{H_p}{4}$ & $J-\frac{H_p}{4}$ &
  $\left(\begin{array}{ll} 2 & 0 \\ 1^- & 1 \end{array}\right)$ & $J$ &
  0 & $J$ & 0 & 0 \\
  $\left(\begin{array}{ll} 0 & 2 \\ 1^+ & 1 \end{array}\right)$ & $J$ &
  0 & $J$ & 0 & 0 &
  $\left(\begin{array}{ll} 0 & 2 \\ 1^- & 1 \end{array}\right)$ & $J$ &
  0 & 0 & 0 & $J$ \\
  $\left(\begin{array}{ll} 1 & 1 \\ 2^+ & 0 \end{array}\right)$ & 0 &
  0 & 0 & 0 & 0 &
  $\left(\begin{array}{ll} 1 & 1 \\ 2^- & 0 \end{array}\right)$ & $J$ &
  0 & 0 & $B'$ & $A'$ \\
  $\left(\begin{array}{ll} 1 & 1 \\ 0^+ & 2 \end{array}\right)$ & $J$ &
  0 & 0 & $A$ & $B$ &
  $\left(\begin{array}{ll} 1 & 1 \\ 0^- & 2 \end{array}\right)$ & $0$ &
  0 & 0 & 0 & 0 
\end{tabular}
\end{ruledtabular}
\end{table*}

\begin{table*}
\caption{
  A $\pm 2$ worm solution
  to the detailed balance equation for the $S=1$
  antiferromagnetic Heisenberg model.
  Applies only if $0\le H_p\le 2J$.
}
\label{tab:PMTwoWeightForAFH}
\begin{ruledtabular}
\begin{tabular}{cc|cccc||cc|cccc}
  $\Sigma$ & $W(\Sigma)$ & $\downarrow$ & $\uparrow$ & $\nearrow$ & $\rightarrow$ &
  $\Sigma$ & $W(\Sigma)$ & $\downarrow$ & $\uparrow$ & $\nearrow$ & $\rightarrow$ \\
\hline
  $\left(\begin{array}{ll} 2 & 2 \\ 2^{++} & 2 \end{array}\right)$ & $0$ &
  0 & 0 & 0 & 0 &
  $\left(\begin{array}{ll} 2 & 2 \\ 2^{{-}{-}} & 2 \end{array}\right)$ & $2H_p$ &
  0 & $2H_p$ & 0 & 0 \\
  $\left(\begin{array}{ll} 2 & 1 \\ 2^{++} & 1 \end{array}\right)$ & $0$ &
  0 & 0 & 0 & 0 &
  $\left(\begin{array}{ll} 2 & 1 \\ 2^{--} & 1 \end{array}\right)$ & $J+\frac{3H_p}{2}$ &
  $H_p$ & $J+\frac{H_p}{2}$ & 0 & 0 \\
  $\left(\begin{array}{ll} 2 & 0 \\ 2^{++} & 0 \end{array}\right)$ & $0$ &
  0 & 0 & 0 & 0 &
  $\left(\begin{array}{ll} 2 & 0 \\ 2^{--} & 0 \end{array}\right)$ & $2J+H_p$ &
  $2J+H_p$ & 0 & 0 & 0 \\
  $\left(\begin{array}{ll} 0 & 1 \\ 0^{++} & 1 \end{array}\right)$ & $J+\frac{H_p}{2}$ &
  0 & $J+\frac{H_p}{2}$ & 0 & 0 &
  $\left(\begin{array}{ll} 0 & 1 \\ 0^{--} & 1 \end{array}\right)$ & $0$ &
  0 & 0 & 0 & 0 \\
  $\left(\begin{array}{ll} 0 & 2 \\ 0^{++} & 2 \end{array}\right)$ & $2J+H_p$ &
  $2J-H_p$ & $2H_p$ & 0& 0 &
  $\left(\begin{array}{ll} 0 & 2 \\ 0^{--} & 2 \end{array}\right)$ & $0$ &
  0 & 0 & 0 & 0 \\
  $\left(\begin{array}{ll} 1 & 2 \\ 2^{++} & 1 \end{array}\right)$ & $0$ &
  0 & 0 & 0 & 0 &
  $\left(\begin{array}{ll} 1 & 2 \\ 2^{--} & 1 \end{array}\right)$ & $J$ &
  0 & 0 & $J$ & 0 \\
  $\left(\begin{array}{ll} 1 & 0 \\ 0^{++} & 1 \end{array}\right)$ & $J$ &
  0 & 0 & $J$ & 0 &
  $\left(\begin{array}{ll} 1 & 0 \\ 0^{--} & 1 \end{array}\right)$ & $0$ &
  0 & 0 & 0 & 0 \\
  $\left(\begin{array}{ll} 1 & 1 \\ 2^{++} & 0 \end{array}\right)$ & $0$ &
  0 & 0 & 0 & 0 &
  $\left(\begin{array}{ll} 1 & 1 \\ 2^{--} & 0 \end{array}\right)$ & $J$ &
  0 & 0 & 0 & $J$ \\
  $\left(\begin{array}{ll} 1 & 1 \\ 0^{++} & 2 \end{array}\right)$ & $J$ &
  0 & 0 & 0 & $J$ &
  $\left(\begin{array}{ll} 1 & 1 \\ 0^{--} & 2 \end{array}\right)$ & $0$ &
  0 & 0 & 0 & 0
\end{tabular}
\end{ruledtabular}
\end{table*}

In addition to the worms changing spin values by 1,
we can introduce worms that change spin values by an arbitrary amount.
For $S=1$, we can introduce $\pm 2$ worms.
In the examples presented in \Sec{Performance}, we used both $\pm 1$ worms
and $\pm 2$ ones.
When a pair of worms are created, the type of worm is chosen
with equal probability from all possible ones.
For instance, when the initial spin state is $l=2$ at the point 
chosen for the pair creation,  a $-1$ or $-2$ worm is possible.
Each one of them is chosen with probability $1/2$.
The extended weight for $\pm 2$ worms are listed in \Tab{PMTwoWeightForAFH}.

In fact, only the $\pm 1$ worms are necessary for making the
algorithm ergodic and for satisfying the detailed balance.
Although it is likely that the $\pm 2$ worms are useful for improving the 
efficiency of the algorithm for more complicated models,
we have not encountered such a case yet.


\end{document}